\documentclass[prl,aps,twocolumn,showpacs]{revtex4}

\usepackage{epsfig} 

\begin{document}

\title{NMR evidence of strong-correlated superconductivity in LiFeAs: tuning toward an SDW ordering}

\author{L. Ma ${^1}$}
\author{J. Zhang $^{1}$}
\author{G. F. Chen $^{1}$}
\author{Weiqiang Yu $^{1}$}
\email{wqyu_phy@ruc.edu.cn} 

\affiliation{
$^1$Department of Physics, Renmin University of China, Beijing 100872, China\\
}
\date{\today}

\begin{abstract}

In this letter, we reported the results of NMR study on LiFeAs single crystals. We find a strong evidence of the 
low temperature spin fluctuations; by changing sample preparation conditions, the system can be tuned toward an 
spin-density-wave (SDW) quantum-critical point. The detection of an interstitial Li(2) ion, possibly locating in 
the tetrahedral hole, suggests that the off-stoichiometry and/or lattice defect can probably account for the 
absence of the SDW ordering in LiFeAs. These facts show that LiFeAs is a strongly 
correlated system and the superconductivity is likely originated from the SDW fluctuations. 

\end{abstract}

\maketitle

The interplay of magnetism and superconductivity is one of the dominant 
themes in the study of unconventional superconductors, such as high-T$_c$ cuprates, organic 
superconductors and heavy fermions, where 
the magnetic fluctuations are crucial to the superconductivity in general \cite{Tranquada1995, Mathur, Jerome1980}. 
This subject has also been extensively studied both experimentally and theoretically 
in the recent discovered iron pnictides \cite{Hosono_Jacs_130_3296}, where high-temperature superconductivity is achieved by 
suppressing a competing spin-density-wave (SDW) state upon chemical doping or pressure \cite{Chen_PRL_100_247002, Rotter_PRL_101_107006, 
Dong2008EPL}). Here the superconductivity emerging in proximity to a spin-density-wave (SDW) quantum critical point (QCP), as well as the persisting 
spin fluctuations shown above $T_C$ \cite{Ning_PRL_104}, support strongly that superconducting pairing is mediated by spin fluctuations.

However, in an iron pnictide LiFeAs, bulk superconductivity up to 18 K, instead of long-range antiferromagnetism (AFM), is found in the ground state 
without nominal doping \cite{Jin_CQ_LiFeAs,Pitcher_CC,Chu_PRB}. Angle-resolved photoemission spectroscopy (ARPES) studies \cite{Borisenko_ARPES, 
Kordyuk_LiFeAs_arpes2, Inosov_PRL} do not see evidence of spin fluctuations. In particular, the superconducting gap seems to be a single isotropic gap 
with a moderate amplitude \cite{Borisenko_ARPES, Kordyuk_LiFeAs_arpes2, Inosov_PRL}, in contrast to the multiple gaps in other iron pnictides which is 
likely originated from Fermi surface nesting and spin fluctuations \cite{Ding_83_47001, Mazin_PRL_101_057003, Kuroki_PRL_101_087004, 
Wang_PRL_102_047005, Cvetkovic_EPL_85_37002, Yao_NJP}. The $\mu$SR studies show that LiFeAs has a different Uemura relation with other pnictide 
superconductors \cite{Pratt_musr}. These facts lead to an everlasting proposal that LiFeAs is a conventional superconductor, rather than a strongly 
correlated superconductor.

Theoretically, the local density approximation (LDA) calculations indicate that LiFeAs has a similar band structure with LaFeAsO, BaFe$_2$As$_2$ and 
NaFeAs, and therefore a similar magnetic ordering in the undoped and a universal origin of superconductivity in the doped materials are expected among 
all compounds \cite{Singh_PRB_78_094511, Nekrasov, Liu_CM, Imada}. Indeed, particularly for the sister compound NaFeAs with the same 111 structure, 
the magnetism \cite{LiSL_NaFeAs, Yu_CM} and superconducting properties \cite{Parker_CM_09090417, XiaCoP} (upon doping) are very similar to the 1111 
(such as $R$FeAsO$_{1-x}$F$_x$ \cite{Chen_Nature_453_761, Chen_PRL_100_247002, Ren_CPL}) and the 122 (such as Ba$_{1-x}$K$_x$Fe$_2$As$_2$  and 
Ba(Fe$_{1-x}$Co$_x$)$_2$As$_2$ \cite{Rotter_PRL_101_107006, Chen_EPL_85_17006}) families. It is conjectured that the absence of AFM in LiFeAs is 
probably caused by lithium deficiency \cite{Pitcher_CC}. However, the lacking of evidences for chemical non-stoichiometry, and the absence of 
Curie-Weiss-like low-temperature SDW spin fluctuations from NMR \cite{Jelic_LiFeAs_NMR, LiZ_JPSJ}, different from the results in 
Ba(Fe$_{1-x}$Co$_x$)$_2$As$_2$ \cite{Ning_PRL_104}, do not seem to support this scenario.       

Therefore the study of whether LiFeAs is a strongly correlated superconductor is certainly important for understanding the correlation among the band 
structure, the magnetism and the mechanism of superconductivity of the high $T_c$ pnictides. In order to resolve this problem, we performed NMR 
studies on LiFeAs single crystals. We first searched for possible SDW order and SDW fluctuations in our high-quality single crystals. For the first 
time, we found evidences of anisotropic spin fluctuations, which can be tuned toward an SDW quantum critical point. We further show spectral evidence 
that the absence of antiferromagnetism in LiFeAs is likely caused by the doping and/or the scattering effect from an additional Li(2) site. 

\begin{figure}
\includegraphics[width=9cm, height=6cm]{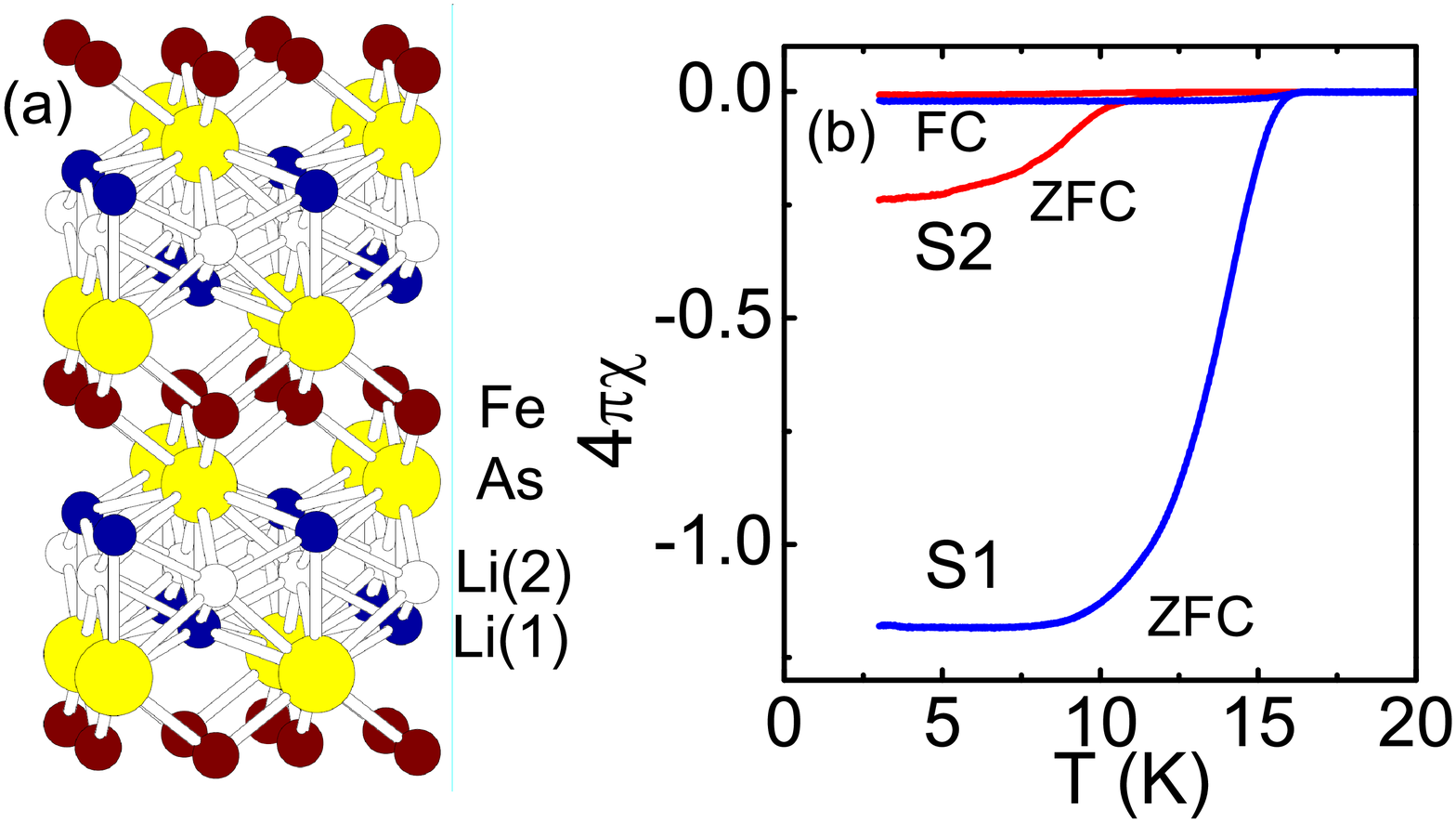}
\caption{\label{mt}(Color online) (a) The proposed crystal structure of Li$_{1+x}$FeAs \cite{Pitcher_CC, Juza}. Additional Lithium ion Li(2) (hollow 
circle) occupies the interstitial site (3/4, 1/4, 1/2), which is right above the Fe site and inside an As$_{4}$ tetrahedron. (b) The dc susceptibility 
of two LiFeAs crystals $S1$ and $S2$ with different growth conditions with zero-field-cooled and field-cooled conditions under 100 Oe field. }
\end{figure}

The single crystals of LiFeAs were grown by self-flux method with two different growth conditions. LiAs was firstly synthesized as precursor
by reacting Li (3N) and As (5N) in Ta tube sealed in evacuated quartz ample and heated at 600 $^{0}$C for 10 hours. Mixtures of LiAs and Fe (4N) 
powder with the composition of Li$_{5}$FeAs$_{5}$ were sealed into Ta tubes under 1.5 atmosphere of argon gas, then the Ta tubes were vacuum sealed 
into quartz tubes. For the sample 1 ($S1$), the ample was heated to 1050 $^{0}$C, held for 24 hours and cooled slowly to 650 $^{0}$C over 200 hours, 
then the furnace is shut down and naturally cooled down to room temperature. The sample 2 ($S2$) was obtained by heating the ample to 1170 $^{0}$C, 
held there for two hours before the temperature was decreased to 870 $^{0}$C within 100 hours, then the ample was taken out and quenched in air. The 
$S1$ has a large superconducting volume and a higher $T_c$ 17 K, whereas the $S2$ was has a smaller superconducting volume and a lower $T_c$ 10 K (see 
Fig.~\ref{mt} (b)). Their electronic and magnetic properties were measured on a Quantum Design physical property measurement system (PPMS) with the 
VSM option provided. 

The NMR crystals were chosen with typical dimension of 3$\times$2$\times$0.1mm$^3$. Both $^{7}$Li ($S=3/2$) and $^{75}$As ($S=3/2$) NMR studies were 
performed with the magnetic field along the $ab-$plane and the $c-$axis. All spectral measurements use the spin-echo technique. The spin-lattice 
relaxation rate is deduced from an inversion-recovery method, and the spin magnetization is fit with the $S=3/2$ nuclear recovery 
$\frac{m(t)}{m(0)}=1-0.1e^{-t/T_1}-0.9e^{-6t/T_1}$ for both $^{7}$Li and $^{75}$As. 

\begin{figure} 
\includegraphics[width=9cm, height=7cm, bb=0 0 800 600]{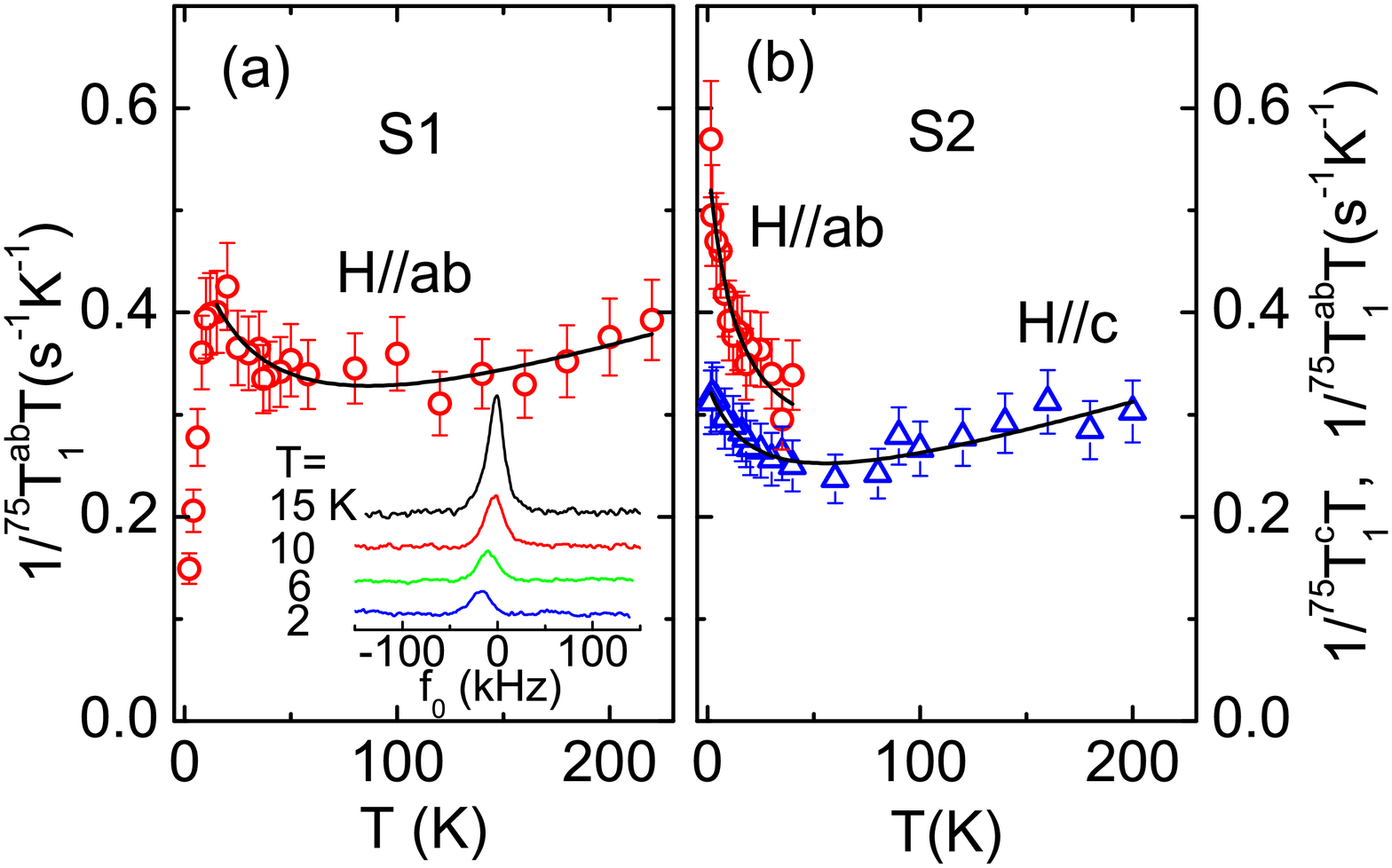}
\caption{\label{invt1t}(Color online) (a) The spin-lattice relaxation rate $1/^{75}T_1T$ of sample $S1$ under a field of 8T applied along the crystal 
$ab$-plane.  The solid line is a fit by a Curie-Weiss term $1/T_1T=A/(T+\Theta)+b+cT$ with $\Theta=30\pm 5$ K. The $^{75}$As NMR spectra at different 
temperatures are shown below. (b) The $1/^{75}T_1T$ of sample $S2$ with field applied along ab-plane and the c-axis. The solid line is a guide to the 
Curie-Weiss fitting  with $\Theta=10\pm 5$ K for field along the ab-plane, and $\Theta=20\pm 5$ K for field along the c-axis. }
\end{figure}

We first study the spin fluctuations through the spin-lattice relaxation rate $1/T_1$ of $^{75}$As in LiFeAs. In Fig.~\ref{invt1t} (a), the 
$1/^{75}T^{ab}_1T$ of the superconducting crystal $S1$ ($T_C\approx$ 17 K) is shown with an 8 T magnetic field along the ab-plane. The Superconducting 
onset is shown by a sharp drop of the $1/^{75}T_1T$ below $T_c$. There is a small upturn while temperature decrease from 50 K down to $T_c$, which 
indicate spin fluctuations. Following Moriya's 2D spin fluctuation theory in a paramagnet, we fit the data with a Curie-Weiss behavior 
$1/T_1T=A/(T+\Theta)+b$ at low temperatures (see Fig.~\ref{invt1t} (a)). Here $A$ is proportional to the electron density of state (DOS) on the Fermi 
surface, and the value of $\Theta$ is correlated with band mass $m^*$, whose sign usually switches from negative to positive if the system is tuned 
from an antiferromagnetic ground state to a quantum disordered paramagnet. The $b$ term is obtained phemenologically, assuming a Korringa (Fermi 
liquid) contribution from the multiple band system. Our fitting parameter $\Theta=30\pm 5$ K is comparable with the optimal-doped 
Ba(Fe$_{1-x}$Co$_x$)$_2$As$_2$ \cite{Ning_PRL_104}, indicating a similar strength of spin fluctuations with other iron pnictide superconductors.
 
To check the sample dependence of spin fluctuations, we investigated the spin-lattice relaxation of sample $S2$ with a less superconducting volume and 
a lower $T_c$. The spin-lattice relaxation rate (SLRR) of the crystal $S2$ is measured with field applied both along the ab-plane ($1/^{75}T^{ab}_1T$) 
and along the c-axis ($1/^{75}T^{c}_1T$) (see Fig.~\ref{invt1t} (b)). Comparing with $S1$, the low-temperature $1/^{75}T^{ab}_1T$ of $S2$ are very 
different. The $1/^{75}T^{ab}_1T$ increase dramatically as temperature drops under a 12 T magnetic field with $\Theta \approx 10 \pm 5$ K, which is 
close to a diverging behavior (i.e., $\Theta \le 0$ K) at finite temperatures. Such behavior is a clear indication that the system is close to a 
magnetic ordering.

Next we discuss the nature of the low-temperature spin fluctuations. As shown in Fig.~\ref{invt1t}, the Curie-Weiss behavior is also seen in 
$1/^{75}T^{c}_1T$ ($\Theta \approx 20 \pm 5$ K), but much weaker than that of the $1/^{75}T^{ab}_1T$. The anisotropy of the spin-lattice relaxation 
rate, defined as $T^{c}_1/T^{ab}_1$, increases as temperature drops, which is well described as signatures of the stripe AFM (or the SDW) correlations 
\cite{Kita_JPSJ_77_114709, Ning_JPSJ_78_013711, Kitagawa_CM}. The observations of the dramatic enhancement and the anisotropy of the low-temperature 
spin-lattice relaxations unambiguously indicate that LiFeAs is a strong-correlated system, which is close to the SDW order. 

This draws a similarity among LiFeAs and other iron pnictides, and suggests that the superconductivity in LiFeAs is also mediated by SDW spin 
fluctuations. Although the long-range magnetic ordering is not achieved so far in our crystals, the small value of $\Theta$ in $S2$ indicates that the 
LiFeAs is tuned toward an SDW quantum critical point. Obviously it is important to find what the tuning parameter is. The sample difference is 
probably related to the chemical off-stoichiometry which contribute to a doping effect, and/or an impurity scattering effect.

\begin{figure}
\includegraphics[width=9cm, height=7cm,bb=0 0 800 600]{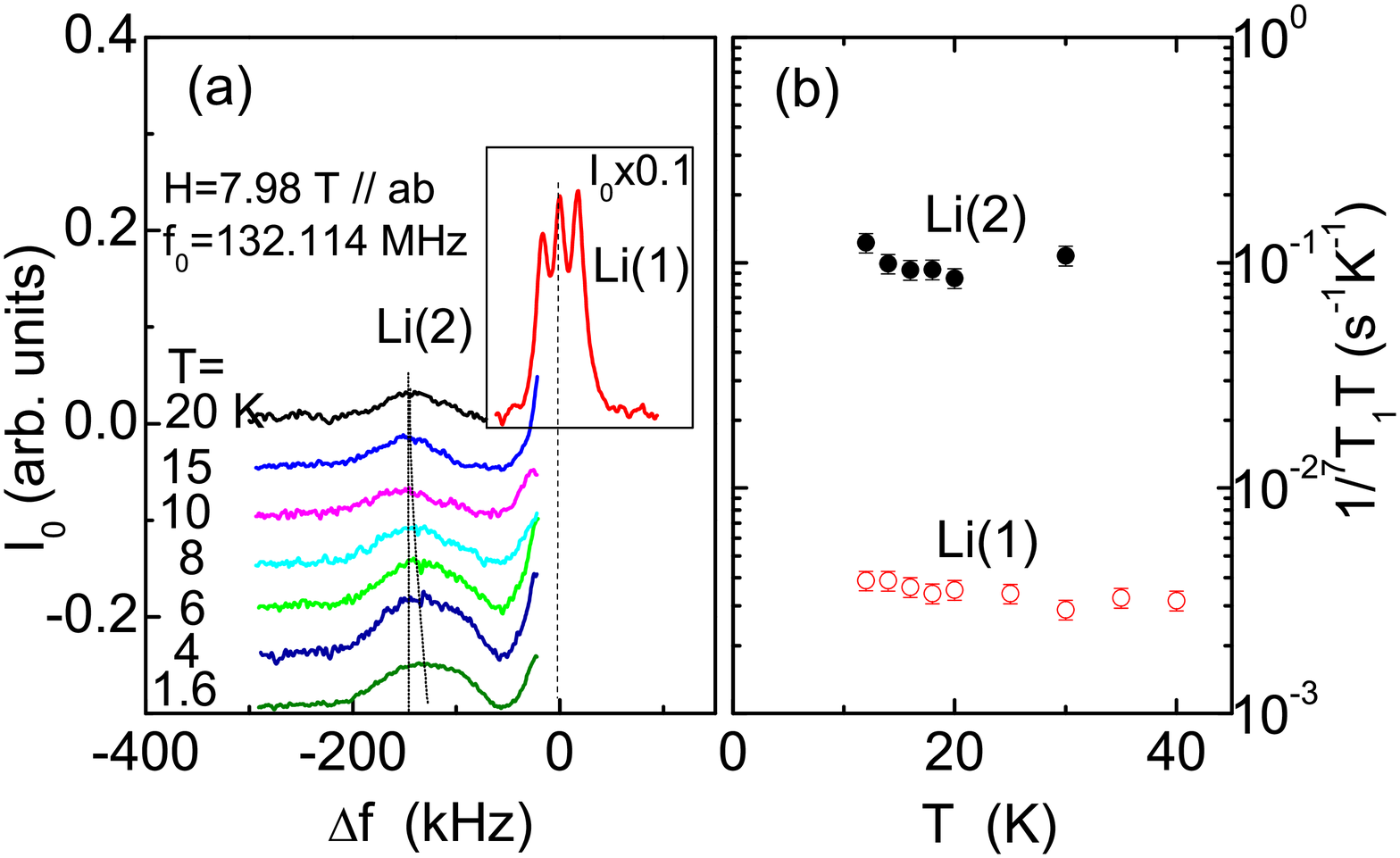}
\caption{\label{spec2}(Color online) (a) The $^7$Li NMR spectrum of a LiFeAs crystal ($S1$) with field along the ab-plane. The spectrum splits into 
two species, Li(1) and Li(2). The Li(2) spectrum below $T_c$ is obtained by a fast recovery method (see text). (b) The comparison of the SLRR between 
Li(1) and Li(2). The $1/^{7}T_1$ on the Li(2) site is thirty time faster, but in the same order with $1/^{75}T_1$ of $^{75}$As.}
\end{figure}

In fact, it is well known that lithium could be reversibly intercalated/inserted into the interstitial sites for many materials due to its high 
mobility and small size, such as in lithium ion batteries. Considering such behavior of lithium and the structure of LiFeAs, we performed $^7$Li 
spectrum studies to search for possible extra Lithium in the lattice. With the magnetic field along the ab-plane, we found a very weak spectra tail on 
the low frequency side of the normal Li spectrum in $S1$ (see Fig.~\ref{spec2}), and we label two sub-species as Li(1) and Li(2). The Li(1) (one 
central transition and two satellites with $^{7}\nu _q (1)\sim$ 0.06 MHz) is identified as normal lithium site as shown in Fig.~\ref{mt}, because of 
its large intensity (multiplied by $0.1$ in the figure), low hyperfine coupling from its small Knight shift ($^{7}K_n\ll 0.005\%$) and slow spin 
lattice relaxation rate.  The Li(2) is on another specific site, because its spin-lattice relaxation follows a single $T_1$ component (data not 
shown). Both Li(1) and Li(2) spectra broaden blow $T_c$, which leads to more spectral overlap in the frequency range. In order to reduce the overlap, 
we use a fast recovery method, where the samples are fed with saturation pulses and then the measurements are taken after a short recovery time $t$ 
($t\approx \frac{1}{6}^7T_1(2) \ll ^7T_1(1)$). Since Li(1) has a much longer $T_1$ than that of Li(2) (see Fig.~\ref{spec2} (b)), the spectral weight 
contribution of Li(1) at the Li(2) frequency is then negligible. 

The Li(2) spectra shift to higher frequency and broaden right below $T_c$, which indicates that Li(2) is intrinsic in a superconducting state. We are 
not able to find the satellite of Li(2) and assign the lattice position of Li(2) directly, possibly because the satellite intensity of Li(2) is too 
low.  According to the spectrum and the spin-lattice relaxation measurements, it is natural to conclude that the Li(2) comes from an interstitial site 
in the lattice (see Fig.~\ref{mt} (a)), which is located just above the Fe site and enclosed in an As$_4$ tetrahedron \cite{Juza}.  Since Li has a 
very small quadruple moment, the quadrupole correction to the center frequency of Li(2) is probably negligible\cite{nqr}. The Knight shift of Li(2) 
with field along the ab-plane is estimated with $^{7}K^{ab}_n\sim -0.1\%$, much larger than that of Li(1). Li(2) is also found with field along the 
$c$-axis, and Knight shift is estimated as $^{7}K^{c}_n\sim -0.05\%$. Early x-ray studies\cite{Juza} on LiFeAs suggest an interstitial Li site located 
above the Fe site and enclosed in an As$_4$ tetrahedron, which seems consistent with our Li(2) data for several reasons: (1) A large hyperfine 
coupling of the Li(2) to the Fe is expected, unlike the cancellation effect of the diagonal hyperfine field on the Li(1) position 
\cite{Kita_JPSJ_77_114709, Yu_CM}. This is consistent with the fast spin-lattice relaxation rate and a large Knight shift of Li(2). (2) The hyperfine 
coupling $A^c_{hf}$ is expected stronger than $A^{ab}_{hf}$, which is consistent with the anisotropic Knight shift of Li(2). (3) $^{57}$Fe also has a 
negative Knight shift from spin contributions with $^{57}K_n\sim -0.1\%$ \cite{Yashima_JPSJ_NMR}, which supports that Li(2) is located at the same 
symmetric position as iron.

It should be noted that, in Li$_{1+x}$MnAs \cite{LiMnAs}, the (3/4, 1/4, 1/2) site  (see Fig.~\ref{mt} (a)) are fully occupied by lithium, whereas 
only a few percent of additional lithium is located in octahedral holes. One can hence speculate that lithium deficiency and/or insertion can occur 
simultaneously in Li$_{1+x}$FeAs. The comparison of the spectral weigh between Li(1) and Li(2) gives a Li(2) concentration of ($6\pm 2)\%$/unit in 
$S1$. Comparison with the sister 111 compound NaFeAs \cite{Parker_CM_09090417}, such carrier level is sufficient to tune the SDW and superconducting 
state. 
 
Now we discuss the cause of property difference between sample $S1$ and $S2$. Li(2) is also found in sample $S2$, estimated with a concentration of 
($9\pm 2)\%$/unit. However, it is unreliable to compare the concentration difference within the error, and we consider several possibilities. If the 
actual concentration of Li(2) is higher in $S1$, a larger electron-doping effect is expected to suppress the SDW and induces the superconductivity, 
which is seen in $S1$. On the other hand, if the actual concentration of Li(2) is higher in $S2$, additional doping effect such as Li(1) deficiency, 
may exist to cancel the Li(2) effect. Then a higher concentration of Li(2) in $S2$ result in the low $T_c$ and bring the system back toward the SDW 
ordering \cite{li22}. In fact, lithium-deficiency on Li(1) has been speculated to coexist with Li(2) doping \cite{LiMnAs,Pitcher_CC, 
Singh_PRB_78_094511}. Besides, disorder could also play an essential role here. Since $S2$ is made with a fast growth condition, disorder scattering 
may suppress the superconductivity and favor the competing SDW ordering. Finally, we should also point out that although Li(2) is coupled to 
superconductivity, we cannot completely rule out the possibility that the Li(2) is from a minor superconducting phase. More work is needed to verify 
the position and the role of Li(2) proposed in LiFeAs.
   
To summarize, we found two independent evidences for strong-correlated superconductivity in LiFeAs. First, evidence of strong spin fluctuations is 
found in the normal state right above $T_c$, which increase as temperature drops. Such effect supports that superconductivity is probably mediated by 
spin fluctuations. In particular, a suppression of $T_c$ with different growth conditions leads to a significant enhancement of anisotropic spin 
fluctuations toward an SDW quantum critical point. Second, our data shows a Li(2) signal with a concentration of ($6\pm 2)\%$/unit in superconducting 
LiFeAs, and the absence of the long range AFM in LiFeAs could be caused by a doping effect. Combining both evidences, our data unifies LiFeAs with the 
1111 and the 122 iron pnictides with the same magnetic origin and the same mechanism of superconductivity. We believe that our results are important 
to understand the mechanism of superconductivity and underline further the importance of magnetic fluctuations for the superconductivity pairing 
observed in iron-based superconductors.
 
The Authors acknowledge Dr. Z. Y. Lu, T. S. Zhao, W. Bao and P. C. Dai for helpful discussions. W.Y. and G.F.C. are supported by the National Basic 
Research Program of China (the 973 project) and the National Natural Science Foundation of China. 


\begin{thebibliography}{42}
\bibitem[{nqr()}]{nqr}
\bibinfo{note}{Assuming a similar electri field gradient on As and Li(2), the
  shift of the center is about 10 kHz with $\mu _0H$= 12T//$ab$ and 0 kHz with
  $\mu _0H$= 12T//$c$}.
\bibitem[{li2()}]{li22}
\bibinfo{note}{The spin-lattice relaxation rate $1/^{75}T_1$ in LiFeAs is close
  to that of the stoichiometric NaFeAs, which suggests that overdoping with
  Li(2) is unlikely.}
\end{thebibliography}

\end{document}